\documentclass{elsart}
\usepackage{graphicx,amssymb,harvard}
\journal{New Astronomy}

\newcommand{\lsim}{\,\lower2truept\hbox{${<\atop\hbox{\raise4truept\hbox{$\sim$}}}$}\,}
\newcommand{\gsim}{\,\lower2truept\hbox{${>\atop\hbox{\raise4truept\hbox{$\sim$}}}$}\,}

\begin{document}

\begin{frontmatter}

\title{On the loss of telemetry data in full-sky surveys from space}

\author[Maino,Pasian]{D.~Maino},
\author[Burigana]{C.~Burigana},
\author[Pasian]{F.~Pasian},

\vskip 0.5truecm

\address[Maino]{Dipartimento di Fisica, Universit\`a di Milano, Via Celoria 16,
I-20131, Milano, Italy}
\address[Burigana]{IASF, Sezione di Bologna, Consiglio Nazionale delle Ricerche, Via Gobetti
101, I-40129, Bologna, Italy}
\address[Pasian]{INAF-Osservatorio
Astronomico, Via G.B.~Tiepolo 11, I-34131, Trieste, Italy}

\footnote{The address to which the proofs have to be sent is: \\
Davide Maino, Dipartimento di Fisica, Universit\`a di Milano,
Via Celoria 16, I-20131, Milano Italy\\
fax: +39-02-5031-7272\\
e-mail: Davide.Maino@mi.infn.it}

\newpage
\begin{abstract}

In this paper we discuss the issue of loosing telemetry (TM) data
due to different reasons ({\em e.g.} spacecraft-ground transmissions)
while performing a full-sky survey with space-borne
instrumentation. This is a particularly important issue considering the
current and future space missions (like {\sc Planck}
from ESA and $MAP$ from NASA) operating from an orbit far from Earth 
with short periods of visibility from ground stations.
We consider, as a working case, the Low Frequency Instrument 
(LFI) on-board the {\sc Planck} satellite
albeit the approach developed here can be easily applied to any
kind of experiment that makes use of an observing (scanning)
strategy which assumes repeated pointings of the same region of
the sky on different time scales. 
The issue is addressed by means of a Monte Carlo approach.
Our analysis clearly shows that, under quite general conditions,
it is better to cover the sky more times with a lower 
fraction of TM retained than less times with a higher guaranteed TM fraction. 
In the case of {\sc Planck}, an extension
of mission time to allow a third sky coverage with 95\% of the 
total TM guaranteed provides a significant reduction 
of the probability to loose scientific information with respect 
to an increase of the total guaranteed TM to 98\%  
with the two nominal sky coverages.

\end{abstract}

\begin{keyword}
Astronomical and space-research instrumentation \sep  Astronomical
observations: Radio, microwave, and submillimeter
\PACS 95.55.n \sep 95.85.Bh \sep 96.30.Ys
\end{keyword}

\end{frontmatter}

\newpage

\section{Introduction}

An increasing number of space missions of astrophysical interest
are avoiding orbits around the Earth, to improve their
environmental conditions. In particular, orbits far from the Earth around the
Lagrangian point of the Earth-Sun system ($L_2$) are currently
being selected especially by infrared and microwave missions,
like $MAP$ by NASA (Bennett et al. 1996) and {\sc Planck}
(Tauber 2000), Herschel (Pilbratt 2000) and GAIA (Perryman 2001) by ESA.
While this solution is often
essential for the successful scientific return of the missions,
non-trivial practical problems need to be solved; among these, the
visibility of the spacecraft from the ground station. 
If the spacecraft is not visible all the time, it needs
to have some built-in autonomy to perform its functions independently
from ground control.

In particular, both house-keeping and scientific data need to be
stored on-board, to be subsequently down-linked to the ground
station during the next period of visibility. The data have to be
``safely" transmitted to Earth, with minimal loss of data during
the communications period, due to the high cost of each
telemetry (TM) packet for space missions and to the wealth of
scientific information encoded in each packet. 
However 
some information will be eventually lost, since the
cost of guaranteeing completely faultless communications and
ground systems would be unbearable, if ever feasible. Therefore,
great attention has to be devoted to assess the total amount of
TM that it is possible to loose without affecting the
scientific return of the considered space mission.

In this paper we want to address this issue and we 
adopt a Monte-Carlo (MC) approach to take more easily and
faithfully into account the properties of the observing
strategy of the experiment under consideration. As a
working case we consider the impact of TM losses for the
{\sc Planck} Low Frequency Instrument (LFI, see
Mandolesi et~al. 1998), designed to map the whole sky in 
temperature and polarization at frequencies 
between 30 and 100~GHz and observe the CMB anisotropy 
with an angular (FWHM) resolution 
from $\simeq 33'$ to $10'$ and a sensitivity per (FWHM$^2$)
resolution element from $\simeq 5$ to 13~$\mu$K in 
the measure of the antenna temperature fluctuations 
($\simeq \sqrt{2}$ worst in the measure of 
fluctuations of the Stokes polarization parameters $Q$ and $U$).
It is, however, worth to note that the formalism and approach developed
here are quite general and applicable in practice to any kind
of experiment with redundant observing strategy (like $MAP$), 
{\em i.e.} where the same sky region is observed on several 
different time scales.
For the specific working case adopted a total number of 100,000 
simulations representing real cases of
lost TM have been considered and analyzed in terms of probability
of not observing sky regions and of dimension of unobserved regions.

\section{The Monte Carlo approach}

Our approach works once details on the observing
strategy of the mission under consideration are available and
properly coded.

%



In our working case
the orbit selected for the {\sc Planck} satellite is a tight
Lissajous orbit around the $L_2$ Lagrangian point of the
Sun-Earth system. The spacecraft spins
at $\sim$ 1 rpm and, in the simplest scanning strategy, the spin
axis is kept on the anti-solar direction at constant solar aspect
angle by a re-pointing of 2.5$'$ every hour. 
The two intruments (LFI and the High Frequency Instrument, 
see Puget et~al. 1998)
on the focal plane of an Aplanatic 
telescope of 1.5 meter aperture have a field of view at $\sim 85^\circ$
from the spin-axis direction. 
They therefore trace large circles in the sky and the 1-hour averaged circle
is the basic {\sc Planck} scan circle.
In the nominal 14 months mission $\sim$~10,200 basic circles
will be considered, covering twice nearly the whole sky, $\sim$~5,100 
circles for each sky coverage.



Data continuously acquired are packed into TM packets and 
sent to a single ground station (located in Perth - Australia) 
during the connection period (2--3 hours a day).
In case of failure of communications with the ground station
data can be stored on-board for a maximum amount of 48 hours of data.
After this period data are progressively deleted and lost.

As for other missions, at least 95\% of the total TM 
is guaranteed to be finally available 
for further analysis.
Higher percentages of received TM, for example up to 98\%,
may require another operating ground antenna,
and/or have other large additional costs. 
We observe that loosing the 
5\% (2\%) of the $\sim$~5,100 scan circles of a single sky 
coverage means to lose $\sim$~10 (4) 24-hour-TM-blocks,
corresponding to a set of unobserved ``stripes'' with a global width 
of $\sim$~10$^\circ$ (4$^\circ$) at low ecliptic latitudes. 
It is therefore of paramount importance to evaluate the impact
of the lost TM on the effective sky coverage.

In the specific case of {\sc Planck}-LFI, the antenna beams corresponding to
the various feed horns arranged in the focal plane 
are located on a ring subtending an angular radius of about $3^{\circ}$ on the 
telescope field of view about the telescope optical axis.
The focal plane arrangement of the feed horns at different frequencies
shows potentially 
dangerous situations for the 30~GHz channels (only 2 placed along the same scan direction)
and the 70~GHz (placed along a small arc with an extension of only $\sim 1$~degree
in the direction orthogonal to the scan direction).
In these cases there is no possibility to compensate for the loss of a given sky area
with retained TM observed by other detectors at the same frequency. This is for example
the case of the 100~GHz channels, that span an angle of about 3-4 degrees: sky is effectively
lost only if it is not possible to communicate with the satellite for more than 3 days, or if
the data stored on-board are not downlinked in time. The first is a very
unlikely case, while to cope with the second, an appropriate downlinking strategy shall
be devised.

It is worth to mention that a similar situation is valid also
for, {\em e.g.},  $MAP$ and GAIA. However to properly address the
issue of loosing TM packets for these missions, details on their
observing strategy as well as on-board data storage capabilities 
are required. 


\subsection{Simulations}

When coding the properties of the {\sc Planck} observing
strategy, we made some simplifying assumptions for the only
purpose of computing the percentage of lost TM:
\begin{itemize}
\item each sky coverage ($\simeq$ 7 months long) is composed of an
integer number of scan circles;
\item the number of scan circles is the same for each sky coverage;
\item scan circles from subsequent sky coverages overlap exactly;
\item TM is lost in chunks 48 or 24 hours long (the first is the total
amount of data that can be stored on-board and implies having lost
2 or 1 complete days of data).
\end{itemize}
Furthermore, the pointing stability of the
spacecraft \cite{buri01} assures that the real situation will be
not much different from the case considered here.

Given the mission duration and percentage of lost TM  
we randomly extract from the full TM
stream the lost scan circles. We then overlap the different sky
coverages to form a single TM stream that refers to the whole sky.
In this stream we consider the total number of scan circles lost,
their mean and their maximum dimension.

\begin{figure}
\caption{Schematic representation of the MC approach used here to
simulate lost TM. The grey blocks represent portions of TM that
is lost during the mission. The three sky coverages exactly overlap to obtain the
final result in which only one block is effectively lost. Of course blocks that are
``grey" in at least one sky coverage have intrinsically lower sensitivity in the final
result.}
\begin{center}
\includegraphics[width=12cm]{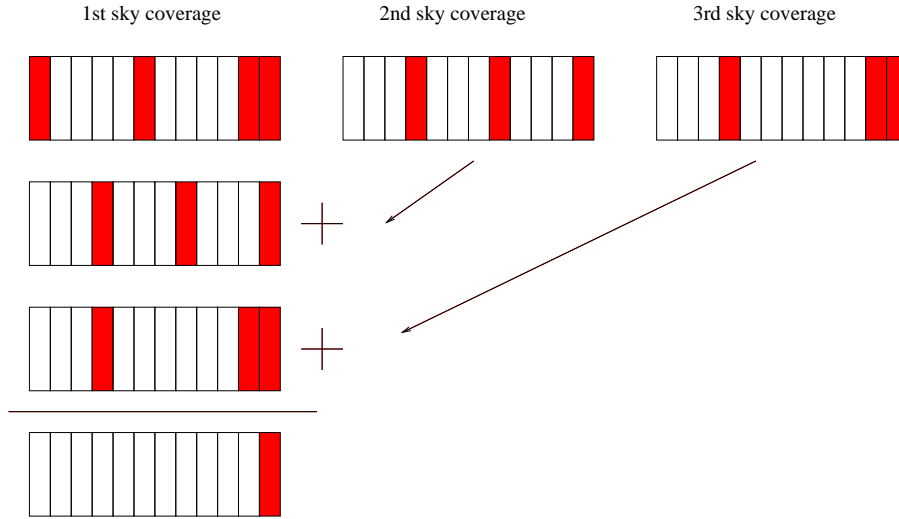}
\end{center}
\end{figure}

We ran over 100,000 MC simulations to derive the probability
distribution function of the lost TM. We assume two different
mission durations implying 2 and 3 complete sky coverages.
Each single sky coverage is composed of a total of
5,100 individual 1 hour scan circles (corresponding to $\simeq$ 7 months).
Two total amounts of lost TM are considered: 5 and 2\%. 

In Table~I we report results from our simulations respectively for 48 and 24 hours
blocks of lost TM: $P(0)$ is the percentage of simulations with no loss
of scan circles at all after coadding a given number of
sky coverages (SC) while $P(\ge 1)$ is the percentage of simulations for which
at least one scan circle is lost at the end of the coadding procedure.
The other two columns report the mean and maximum number of scan circles lost 
in our 100,000 simulations.
\begin{table}[h]
\caption{Results from 100,000 simulations with realistic loss of TM}
\begin{tabular}{lccccc}
\hline
\ TM Lost & TM Block & $P(0)$ & $P(\ge 1)$ & Mean \# & Max \# \\
\hline
\ 5\% - 2 SC & 48 & 66.00 & 34.00 & 28.50$\pm$17.16 & 118\\
\            & 24 & 38.23 & 61.77 & 18.19$\pm$11.64 & 78\\
\ 5\% - 3 SC & 48 & 97.52 & 2.48  & 16.25$\pm$11.41 &  54\\
\            & 24 & 94.35 & 5.65  &  8.64$\pm$5.89  & 37\\
\ 2\% - 2 SC & 48 & 94.64 & 5.36  & 24.05$\pm$13.97 &  76\\\
\            & 24 & 87.45 & 12.55 & 12.85$\pm$7.29  & 46\\
\ 2\% - 3 SC & 48 & 99.90 & 0.10  & 17.19$\pm$10.86 &  48\\
\            & 24 & 99.68 & 0.32  &  7.56$\pm$5.06  & 20\\
\hline
\end{tabular}
\end{table}

While it is of course intuitive that a higher number of sky coverages decreases the
probability of exact and partial overlap of portions of lost TM in each sky coverage,
Table~I clearly quantifies that the case with 5\% of lost
TM and with 3 sky coverages is considerable better than the case with only 2\% of lost
TM and 2 sky coverages. 

In Figure~2 and Figure~3 we report the distribution of the number of scan circles lost
after coadding sky coverages together, when considering loosing TM in 24 and 48 hours
blocks respectively.
\begin{figure}
\caption{Distribution of number of scan circles lost - 24 hours sets}
\begin{center}
\includegraphics[width=12cm]{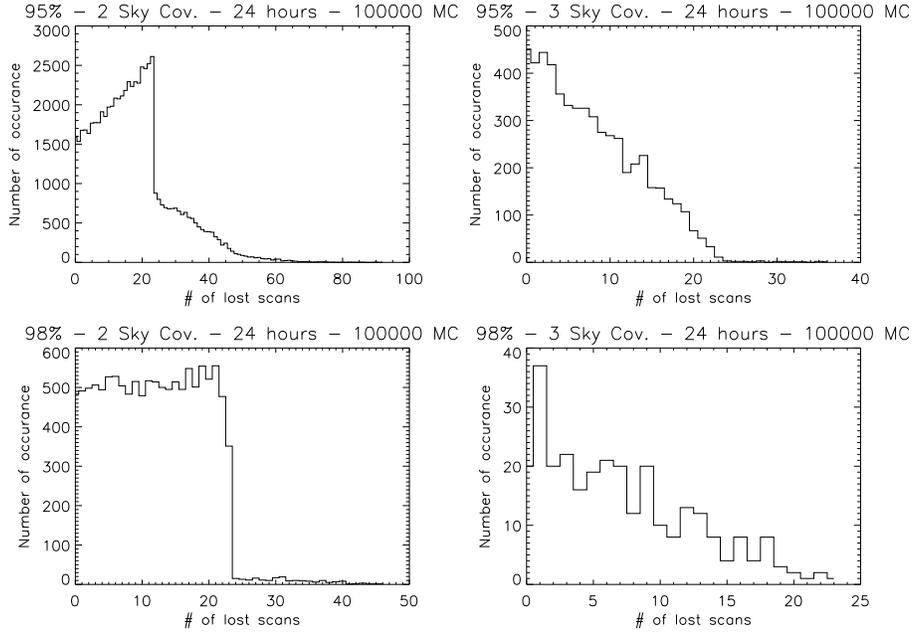}
\end{center}
\end{figure}
\begin{figure}
\caption{Distribution of number of scan circles lost - 48 hours sets}
\begin{center}
\includegraphics[width=12cm]{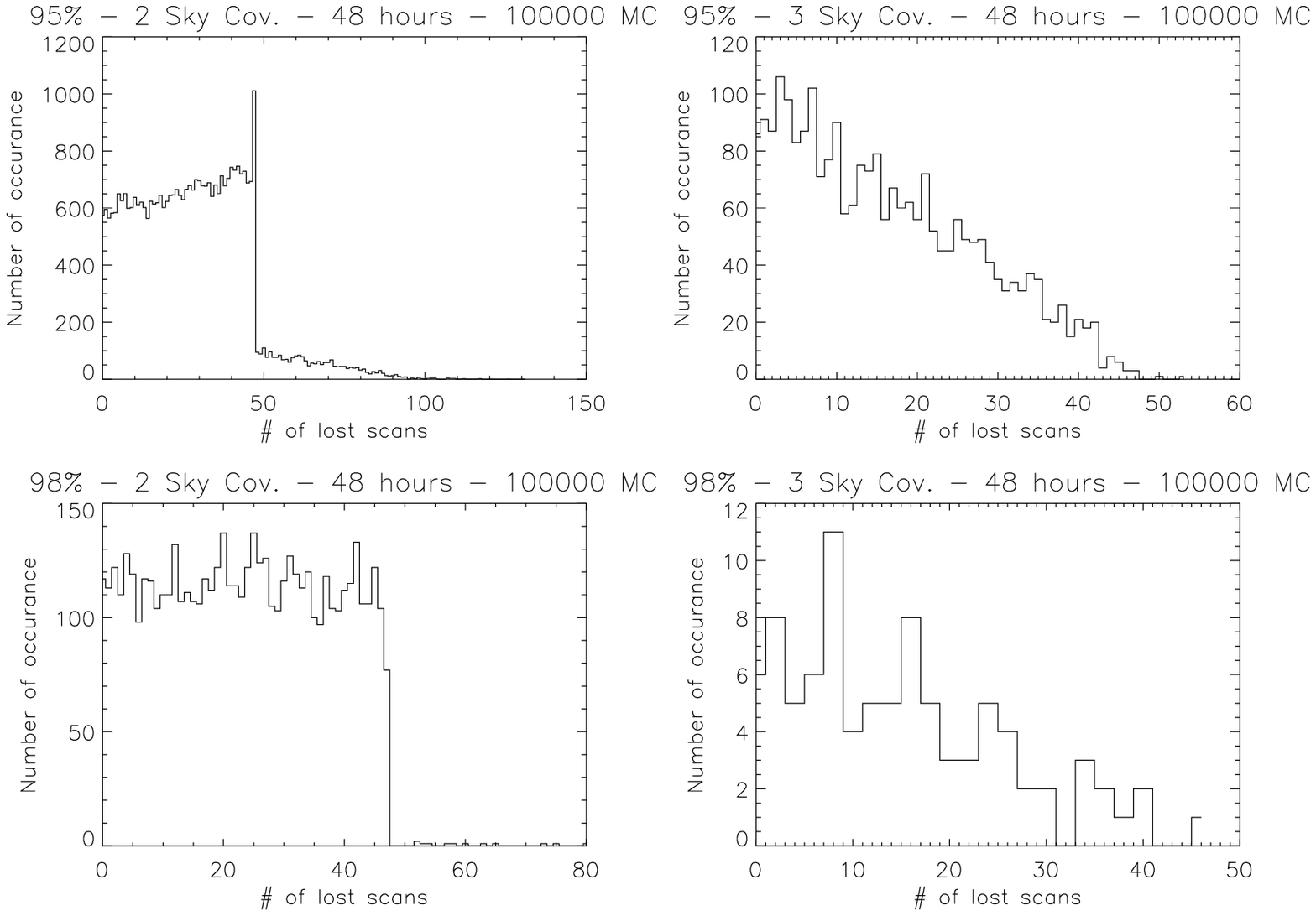}
\end{center}
\end{figure}

One interesting feature of these plots is that the cases with 3 sky coverages show a rapidly
decreasing distribution with increasing number of scan circles lost while the cases with
only 2 sky coverages have a much flatter distribution up to the dimension of the
considered TM block. This means that in the latter cases there
is almost the same probability of loosing a single scan circle or loosing 24 (48) or
more scan circles. There is a clear sharp cut-off 
in these distribution functions
at 24 (48) for the 24 (48) hours sets, respectively. Furthermore
from a detailed study of the distribution of contiguous lost scan circles, we find that 
the probability of loosing sets of
contiguous 24 (48) hours 
of scan circle 
is below 1\% (0.1\%) considering 5\% of lost TM.
These numbers fall futher down when considering 2\% of lost TM.

\subsection{Sky Fraction Lost}

Another important aspect is the dimension of the lost TM
projected into the sky when coadding the whole set of TM onto a sky map.
This final step is somewhat dependent on the final resolution of the map to be
created, {\em i.e.}
the pixel size chosen for the map that is related to the beam width of the instrument
collecting data. 
To evaluate the total sky fraction effectively lost, we first evaluate
the probability distribution of the lost scan circles dimension.
This goes, for the case of {\sc Planck}, 
from the elementary dimension
of a single ring of 2.5$'$ up to a maximum value. Figures~4 and 5 show this probability
distribution for the cases considered here.
\begin{figure}
\caption{Distribution of lost scan circles dimensions - 24 hours sets}
\begin{center}
\vspace{0.2cm}
\includegraphics[width=12cm]{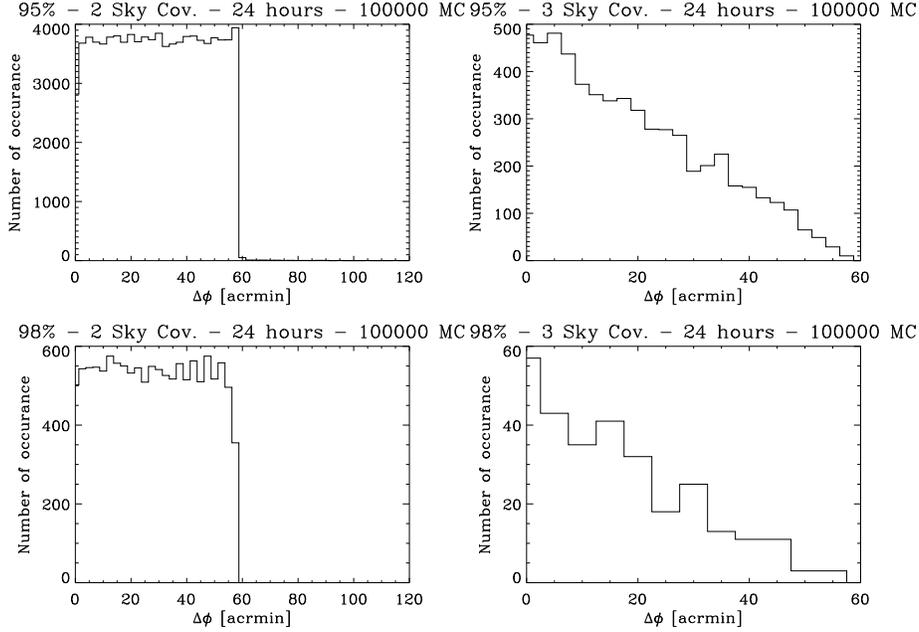}
\end{center}
\end{figure}

\begin{figure}
\caption{Distribution of lost scan circles dimensions - 48 hours sets}
\begin{center}
\vspace{0.2cm}
\includegraphics[width=12cm]{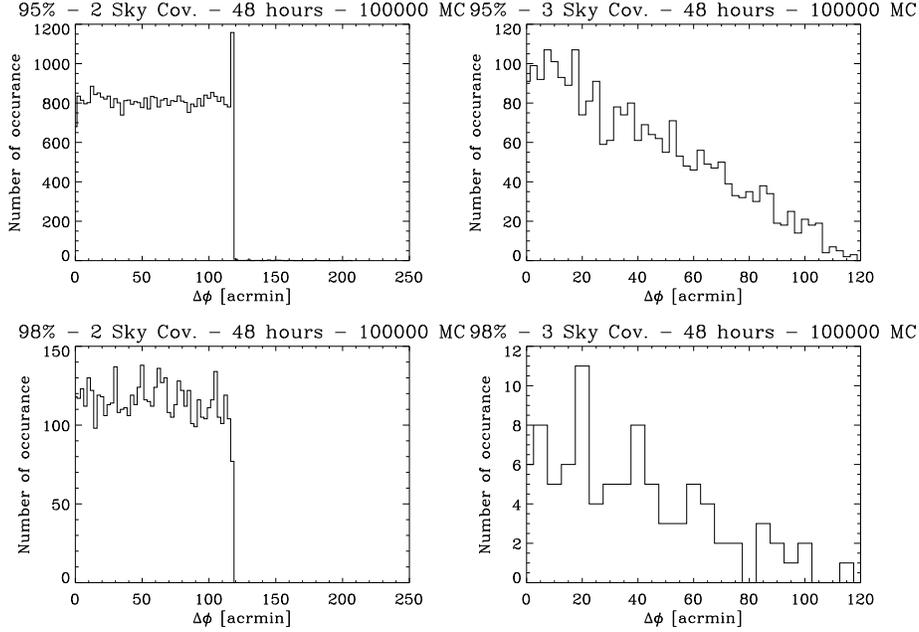}
\end{center}
\end{figure}

As already seen for the distribution of number of scan circles lost, also in the
distribution of dimensions we observe a flat distribution for the cases with 2 sky
coverages while a rapidly decreasing distribution is present for 3 sky coverages.
The same sharp cut-off observed in Figures~2 and 3 is clearly present in
Figures~4 and 5 only for the
5\% of lost TM and 2 sky coverages. The probability of loosing set of contiguous 
scan circles larger than the dimension of 24 (48) scan circles is practically zero.

It is now possible to evaluate
the fraction of the sky that is left unobserved. 
This of course can be done only
in a statistical sense\footnote{We will derive information of
the total amount of sky effectively
lost but we do not know how this fraction is distributed on the sky.}.

\begin{figure}
\caption{The equatorial displacement $\Delta \phi$ is the dimension of a given
set of contiguous scan circles lost. The co-latitude $b_{\rm min}$ is the value at which
the distance between the two meridians is comparable with pixel size.}
\begin{center}
\vspace{0.2cm}
\includegraphics[width=4cm]{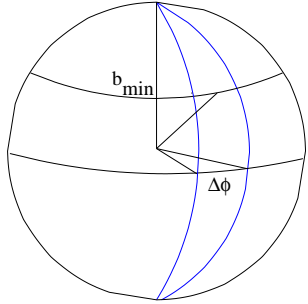}
\end{center}
\end{figure}
In Figure~6, the two meridiands are the edges of the set of scan circles lost
covering an angle $\Delta \phi$ along the equator. The value of $b_{\rm min}$ is derived
from the map pixel size: it is the value of the co-latitude at which the distance
between the edges in the figure is comparable with the pixel size. 
The ``elementary'' lost area is delimited by the edges, the equator and the
isolatitude circle at $b_{\rm min}$. For evident symmetry reasons there
are four of these ``elementary'' areas. The area of a lost scan chunk is then given by:

\begin{equation}
\Omega = 4 \times \int_0^{\Delta \phi} d\phi \int_{b_{\rm min}}^{\pi/2} d\theta {\rm sin}(\theta)
\end{equation}

Knowing the distribution function $N(\Delta \phi)$ (see Figures~4 and 5), 
the total fraction of sky lost can be derived:
\begin{equation}
Area = \sum_i N(\Delta \phi) \Omega_i
\end{equation}
where summation start from $\Delta \phi$ comparable with the pixel size,
typically assumed to be $\sim 1/3$ of the beam FWHM.

In Table~II we report the fraction of sky effectively lost 
for pixel size of 13.7$'$ and 6.8$'$ that are
representative for a map at 30~GHz (FWHM=33.6$'$) and at 70~GHz (FWHM = 22.3$'$)
that are the most critical {\sc Planck}-LFI channels with respect to this issue.
We report here only the case for 48 hours sets of lost TM.

\begin{table}[ht]
\caption{Summary of sky fraction lost for two different pixel sizes - 48 hours set}
\begin{tabular}{lccc}
\hline
\ TM Lost & Pixel size [$'$] & Area [Sq. degress] & Fraction [\%] \\
\hline
5\% - 2 SC & 13.7 & 88.62 & 0.21 \\
5\% - 3 SC & 13.7 &  3.62 & 8.8 $\times 10^{-3}$ \\
2\% - 2 SC & 13.7 & 12.14 & 0.029 \\
2\% - 3 SC & 13.7 &  0.13 & 3.1 $\times 10^{-4}$ \\
5\% - 2 SC &  6.8 & 91.40 & 0.22 \\
5\% - 3 SC &  6.8 &  3.87 & 9.4 $\times 10^{-3}$ \\
2\% - 2 SC &  6.8 & 12.54 & 0.030 \\
2\% - 3 SC &  6.8 &  0.14 & 3.3 $\times 10^{-4}$ \\
\hline
\end{tabular}
\end{table}

Inspection of Table~II shows that the area of lost sky depends on the pixel size of
the map only weakly: when the pixel size decreases by a half the area of the lost sky increases
only by a tiny fraction. 
The improvement represented by the case of 5\% lost
TM with 3 sky coverages with respect to the case with 2\% lost TM but only 2 sky
coverages is quite clear.

\section{Discussion and Conclusions}

We have derived through MC simulations the probability of loosing TM packets for a
space-borne instrument performing a full-sky survey. 


Of course, it is obvious that the best results for a sky coverage in terms of completeness
are obtained when the largest fraction of TM is retained and the number of repeated full
sky observations is increased. However, our analysis clearly shows that it is better to
cover the sky more times with a lower fraction of TM retained than less times with a higher
TM fraction. 

In this respect we note that the 5 year full sky mission GAIA assumes 5 repetitions of 
essentially the same scanning strategy, year by year. Even in the most conservative case
in which a given sky region is observed only one time per year, and considering that: 
{\em i)} the total capacity of the on-board data recorder is about 1 day, {\em ii)} 
one day is also the time scale for the $\sim 1^\circ$ re-pointing of the symmetry axis of GAIA scanning 
strategy, and {\em iii)} the field of view is about $1^\circ \times 1^\circ$ (not far from 
the {\sc Planck} beam size at lower frequencies), we have a
probability to loose a given sky region with 95\% of guaranteed TM is less that
$3\times 10^{-5}$\%. This is another remarkable example in which the increase in the
number of full sky surveys of the mission allows to significantly reduce the
probability of loosing scientific information.

In a space mission, trading off the pergentage of guaranteed TM delivered to the
ground versus number of full sky surveys has an impact in
terms of costs: the setup
needed to guarantee a higher TM fraction may imply more ground stations to follow the
satellite and more ground personnel. Cost-wise, it could be preferable to make an extension
of mission time that implies, {\em e.g.} in the case of {\sc Planck}, only
another seven months of operations. In any case, a careful costs-to-benefits analysis needs to
be carried out. 

\vskip 1truecm

{\bf Acknowledgements}

\noindent
It is a pleasure to thank Floor Van Leeuwen, Michael Perryman and Jos\'e-Luis
Pellon-Bailon for useful discussions on the GAIA scanning strategy. We warmly thank the referee for
constructive comments.

\end{document}